\documentstyle[12pt,aaspp]{article}
\def\Ginga{\hbox{\it Ginga}}
\tighten

\received{1994 August 15}
\revised{ }
\accepted{1995 January 13}
\journalid{VOL}{JOURNAL DATE}
\articleid{Start}{Finish}
\paperid{Manuscript id}
\lefthead{BAND}
\righthead{BATSE GRB Line Search: III. Line Detectability}

\begin{document}
\title{BATSE GAMMA-RAY BURST LINE SEARCH: \\ III. LINE DETECTABILITY}
\author{D. L. Band, L. A. Ford, J. L. Matteson}
\affil{CASS, University of California, San Diego, La Jolla, CA  92093}
\author{M. S. Briggs, W. S. Paciesas, G. N. Pendleton, R. D. Preece}
\affil{University of Alabama at Huntsville, Huntsville, AL 35899}
\author{D. M. Palmer, B. J. Teegarden, B. E. Schaefer}
\affil{NASA/GSFC, Code 661, Greenbelt, MD 20770}

\centerline{\it Received 1994 August 15; accepted 1995 January 13}
\centerline{To appear in {\it The Astrophysical Journal}}

\begin{abstract}
We evaluate the ability of the BATSE spectroscopy detectors to detect the
absorption lines observed by \Ginga\ in gamma ray burst spectra.  We find that
BATSE can detect the 20.6~keV line in the S1 segment of GB870303 with a
detection probability of $\sim 1/4$ in nearly normally incident bursts, with
the probability dropping off to nearly 0 at a burst angle of 50$^\circ$; the
lines at 19.4 and 38.8~keV lines in GB880205 have a high detection probability
in BATSE up to burst angles of $75^\circ$.  In addition, we calculate detection
probabilities for these two line types as a function of signal-to-noise ratio
and burst angle for use in detailed comparisons between BATSE and \Ginga.
Finally, we consider the probability averaged over the sky of detecting a line
feature with the actual array of BATSE detectors on {\it CGRO}.
\end{abstract}
\section{INTRODUCTION}
What are the line detecting capabilities of the Burst and Transient Source
Experiment (BATSE) on board the {\it Compton Gamma Ray Observatory (CGRO)}?  As
described by the first paper in this series (Palmer et al. 1994b), currently no
absorption features have been detected in the spectra observed by BATSE's
spectroscopy detectors (SDs), while two sets of lines were discovered in the
bursts observed by the \Ginga\ gamma ray burst detector. In the second paper of
this series (Band et al. 1994) we presented both Bayesian and ``frequentist''
methodologies for comparing the consistency of the \Ginga\ and BATSE results.
These comparisons require quantitative detection probabilities for lines in
each spectrum accumulated by the two detectors; in the present paper we
calculate these probabilities for the BATSE detectors.  In addition, if lines
are detected in the BATSE bursts, we will then estimate the rate at which lines
occur, which also requires the line detectability probabilities.

The existence of absorption lines is one of the crucial remaining questions
which BATSE can address in the study of GRBs.  One of the key pieces of
evidence supporting the Galactic neutron star origin of GRBs was the series of
reported absorption features in the 15-75~keV band (Konus---Mazets et al. 1981;
{\it HEAO~1}---Hueter 1987; \Ginga---Murakami et al. 1988) interpreted as
cyclotron absorption in a teragauss magnetic field (e.g., Wang et al. 1989);
the only known astrophysical body with such strong fields is a neutron star.
BATSE found that the burst distribution is isotropic yet radially non-uniform
(Meegan et al. 1992), overturning the Galactic disk neutron star paradigm;
whether BATSE also finds absorption lines is the subject of this series of
papers.  If these lines are discovered they will be an important probe of the
burst environment; in particular, it will be difficult (but probably not
impossible) for cosmological models to explain absorption features.  However,
in the absence of any conclusive BATSE detections (Teegarden et al. 1993; Band
et al. 1993a; Palmer et al. 1993, 1994a,b) we have to consider the consistency
between the BATSE and \Ginga\ results (Band et al. 1994), the calculation of
which requires the number of bursts observed by each instrument in which lines
would be detectable.

To properly study the rate at which lines occur in the bursts observed by past
and present missions, we need both a quantitative assessment of the reality of
the claimed lines (as described below) and a list of the spectral parameters
(e.g., the signal-to-noise ratio) which determine line detectability.
Unfortunately, fit parameters were not provided for the line features reported
in $\sim 20$\% of the Konus bursts (Mazets et al. 1981); in addition, the
searched spectra have not been described in quantitative detail. The two line
detections reported among the 21 {\it HEAO~1} bursts (Hueter 1987) are only of
moderate significance (by the methods described below) and few details have
been provided concerning the spectra searched in this data set. However, the
burst data necessary to evaluate both the line candidates and line
detectability in the bursts observed by BATSE and \Ginga\ are available.
Consequently our quantitative comparison is between these detectors.

The absorption features are undoubtedly characterized by distributions of line
parameters.  However, there have been too few definitive line detections to
determine these distributions.  Therefore we use the two \Ginga\ detections as
exemplars of the lines which might be present.  While four sets of lines were
reported, the sets in the S2 segment of GB870303 (Murakami et al. 1988) and in
GB890929 (Yoshida et al. 1991) are not sufficiently significant (by the
detection criteria discussed below) to be considered detections, while the
lines in the S1 segment of GB870303 (Graziani et al. 1992) and in GB880205
(Murakami et al. 1988) are regarded as real.  Therefore we study the
detectability of the single line in the S1 segment of GB870303 (20.6~keV) and
the harmonically spaced lines in GB880205 (19.4 and 38.8~keV) in the BATSE
detectors as functions of the relevant parameters.

In this paper we explore the detectability of absorption lines as a function of
various parameters by the BATSE SDs, both to demonstrate that BATSE can indeed
detect lines similar to those observed by \Ginga, and to derive detection
probabilities which will be useful for future studies.  After describing the
detector (\S 2), we outline the methodology used for this study (\S 3).  The
presentation of our results follows (\S 4).  Finally, we analyze the dependence
of the detectability on various instrumental parameters, and consider the
average detectability of \Ginga-like lines in the BATSE detectors (\S 5).  The
implications of these detection probabilities for BATSE-\Ginga\ consistency
will be discussed later in this series of papers.

\section{THE DETECTORS}

Because more complete descriptions of the BATSE SDs (Fishman et al. 1989, 1995;
Band et al. 1992) are available elsewhere, here we provide a brief overview of
the detectors, emphasizing aspects which are important for understanding their
line detection capabilities.

Each of BATSE's eight modules contains an SD, and therefore every burst is
observed by a number of these detectors; however, not all the SDs from which
burst data are available may have been configured to provide useful spectra
(e.g., as a result of the discriminator and gain settings).  The SDs are simple
scintillation detectors built around a 5" diameter by 3" thick cylindrical
NaI(Tl) crystal with no active shielding.  The detectors are sensitive to gamma
rays entering the side of the crystal above $\sim15$~keV, and thus the maximum
effective area, obtained for sources at an angle of $\sim 30^\circ$ from the
detector normal, is greater than the nominal 127 cm$^{2}$ area of the top of
the crystal at energies where the aluminum case is transparent.  To increase
the response at low energy, the aluminum case has a 3" diameter beryllium
window in the top surface; thus the effective area at $\sim10$~keV is greatest
for sources normal to the detector.

The energy of a photon detected by the SDs is determined by a pulse height
analyzer with 2752 linear channels which are then compressed to 256
quasi-logarithmic channels for transmission to the ground.  The gain of the
photomultiplier tube and the energy at which the lower level discriminator
triggers the pulse height analyzer determine the energy range covered; in
general a spectrum's 256 channels span two energy decades.  Because of the
priority accorded the line search, for most of the mission the gain of six of
the eight SDs has been set to provide spectra as low as is technically
feasible.  As will be discussed below, because of detector-dependent effects,
the low energy cutoff varies by detector.  Unfortunately, after launch an
electronic artifact was discovered which distorts the spectrum for a few
channels just above the spectrum's low energy cutoff, effectively raising this
cutoff.  Although this artifact may be partially mitigated in the future by the
calibration software, the channels of the artifact will probably never be
trusted for more than determination of the continuum below a line candidate
(Band et al. 1992).  However, for the purpose of establishing consistency among
all detectors we do search for features in the channels affected by this
artifact when lines have been found at the same energy in other detectors; for
example, in GB930506 the line candidate at 55~keV in one SD should have
produced a feature just above the artifact in a second SD if it were a true
absorption feature (Ford et al. 1994).

After a burst triggers BATSE, the experiment collects the data necessary to
analyze the event.  In particular, the SDs accumulate a series of 192 spectra
from the four most brightly illuminated modules. The accumulation times are
varied to provide comparable signal-to-noise ratios (SNRs) for each spectrum,
and to cover a period of 4-10 minutes over which the burst may last (it is of
course impossible {\it a priori} to determine when the burst will end); during
analysis on the ground various averages of these spectra are considered.  These
spectra are the basic data which are searched; later papers in this series will
consider how the search is conducted and how all possible averages should be
treated in the statistical analysis.

The mapping of the input photon spectrum into the count spectrum is
traditionally broken into the calibration, which assigns an energy to the
detector channels, and the detector response, which models the detector
efficiency, energy broadening, and processes which distribute counts over a
range of channels.  In the BATSE SD calibration the energy is a monotonic
function of channel number (Band et al. 1992) while the detector response is
written as a matrix equation relating photon and channel energies. Using Monte
Carlo simulations, the response of a BATSE SD was calculated as a function of
the burst angle, the angle between the detector normal and the burst (Pendleton
et al. 1989); the response was assumed to be the same for all detectors and
also independent of azimuth (the angle around the detector normal).  The
response matrix equation is
\begin{equation}
C_i = D_{ij} F_j
\end{equation}
where $C_i$ is the count spectrum, $F_j$ is the binned photon spectrum, and
$D_{ij}$ is the detector response matrix.  The energy channels need not be the
same for $C_i$ and $F_j$, and indeed $F_j$ covers an extended energy range
(particularly on the high side) to model the scattering of photons into the
detectors' nominal energy range.  The BATSE response matrix has been separated
into direct and Earth-scattered components.  The direct component includes
photons which scatter off the spacecraft, while the Earth-scattered components
consist of photons which impinge on the spacecraft from different directions
after scattering off the atmosphere.  The calculations included here do not
include the effects of Earth-scatter.

\section{METHODOLOGY}

We use computer simulations to calculate the detectability of various types of
absorption lines.  A model photon spectrum is convolved with the appropriate
detector response to produce a count spectrum without noise.  By adding Poisson
noise to this simulated count spectrum we create a large number of realizations
which we then fit with continuum and continuum+line models as though they were
observed spectra.  Figure~1 shows simulations of the two \Ginga\ lines as they
might have appeared in the BATSE SDs.  Line significances are calculated from
the fitted parameters.  We are primarily interested in the fraction of the
simulations for which the significance of the features exceeds various
detection thresholds; this fraction is the probability that the feature would
be considered a detection using a given threshold.

We map out the functional dependencies of the line detectability by varying the
components of this procedure.  First, we use response matrices corresponding to
different burst angles.  Second, we model the effect of varying the upper
energy cutoff of the range over which the models were fit.  We fix the low
energy cutoff at 10~keV, a value which applies to some burst-detector data but
is optimistic for most spectra.  Third, we test the sensitivity to the
complexity (i.e., the number of parameters) of the line model.

Fourth, we change the signal-to-noise ratio (SNR), one of the key parameters,
by varying the live time.  The signal is the source flux convolved with the
detector response, while the noise is assumed to result solely from the Poisson
fluctuations of the signal and background counts (we use the background from an
observed burst). The uncertainty in the background determination is not modeled
and thus its contribution to the noise is not included in the SNR. Therefore
the SNR is proportional to the square root of the live time.  For spectra
accumulated over the observed persistence times the SNR in the 25-35~keV band
is $\sim 4.5$ for GB870303 and $\sim37$ for GB880205. Note that varying the SNR
by changing the live time is not exactly equivalent to using different
signal-to-background ratios (SBRs) since the SNR does not vary uniformly across
the spectrum as the SBR is changed: regions of the spectrum where the signal is
much larger than the background will see a smaller change in the SNR than
regions where the background dominates.  The SBR for our GB870303 simulation
varied between 0.04 in the middle of the absorption line to 0.5 at $\sim
150$~keV, while for the GB880205 line the SBR ranged between $\sim 1$ over
$E=10-40$~keV and 4.5 over $E=100-200$~keV. However, simulations where we
varied either the signal strength or the live time both show nearly the same
dependence on SNR; therefore our method for calculating this dependence is
robust.

We fit the simulated spectra with the standard Marquardt-Levenberg algorithm
minimizing $\chi^2$ (Bevington 1969, pp.~232-241; Press et al. 1992,
pp.~678-683).  In brief, the parameters of a model photon spectrum are varied
to minimize the difference---quantified by $\chi^2$---between the calculated
count spectrum (the model photon spectrum folded through the instrument
response) and the observed spectrum.  We define $\chi^2$ with model variances,
requiring additional terms in the gradients used by the Marquardt-Levenberg
algorithm (see the appendix of Ford et al. 1995).

We use the {\it F}-test to determine the line significance.  Assume the
continuum-only fit gives $\chi_1^2$ with $\nu_1$ degrees-of-freedom, while the
continuum+line fit results in $\chi_2^2$ with $\nu_2$.  We define the statistic
$F$ as
\begin{equation}
F = \left({{\chi_1^2-\chi_2^2}\over{\nu_1-\nu_2}}\right)
   \left({{\nu_2}\over{\chi_2^2}}\right) \quad .
\end{equation}
We choose a maximum probability $P(>F)$ as the threshold for a candidate to be
considered a line.  This {\it F}-test probability is defined as the
significance of the feature.  Note that this is only the probability that the
observed line is a fluctuation; it does not consider all the possible
``trials'', the range of parameters (e.g., line centroids, widths, etc.) in
which line-like fluctuations could have occurred.  The true probability that
the line is a fluctuation is the product of the {\it F}-test probability and
the number of trials; therefore we require a very small value of the {\it
F}-test probability to conclude the line is real.

We parameterize burst detectability primarily by the SNR in a number of bands
across the spectrum; the dependence on SNR will be discussed below.  Since
burst continua range in hardness (Band et al. 1993b), the SNR in a band near
the line centroid should be used.  We generally characterize the BATSE spectra
in the 25-35~keV band which is between the lines detected at 20 and 40~keV. We
use only complete channels in these energy ranges, and therefore the actual
energy width varies from burst to burst (the gain and consequently the
energy-channel mapping changes with time).  Since the SNR is proportional to
the square root of the energy range $\Delta E$ over which it is calculated, we
use SNR/$\Delta E^{1/2}$ as the quantity characterizing the signal strength in
each band.

As a consequence of statistical fluctuations, the line significance in the
simulated spectra has a broad distribution, as shown by Figure~2.  The line
detection probability $p($SNR$, \theta)$ is the fraction of the simulations in
which the lines were significant enough to be considered detections.
Graphically, $p($SNR$, \theta)$ is the fraction of the area under the curve in
Figure~2 to the left of the vertical lines indicating two different possible
detection thresholds. Since this fraction is a number based on a finite number
of simulations, there is an associated uncertainty which we approximate as
$\sigma \sim \sqrt{p(1-p)/N_s}$, where $p$ is the calculated probability and
$N_s$ is the number of simulations (usually $N_s=200$).  For $p=0$ or 1 we
calculate $\sigma$ with $p=1/N_s$ or $1-1/N_s$, respectively.  To test this
approximation we ran 12 sets of 200 simulations at two different burst angles
and two different significance thresholds; we found that the actual dispersion
ranged from 1 to 1.4 times $\sigma$.

The line models which are used to characterize the \Ginga\ observations are
either additive or multiplicative Gaussians (see eqns. [3] and [4] below).
Consequently, lines are parameterized by three quantities:  an intensity (e.g.,
an equivalent width), a line centroid and an intrinsic width.  At times the
line models can be simplified, and fewer than three parameters per line are
required.  Models with fewer parameters may result in more significant fits
since both the $F$-test and Bayesian odds compare the improvement in the fit to
the number of added parameters.  Two lines were observed from GB880205 with
harmonically spaced energy centroids at 19.4 and 38.8~keV. Thus the spectra
from this burst can be fit by a line model with only five parameters since the
energy of the second line can be fixed at twice the energy of the first.  The
intrinsic line width is often smaller than, or comparable to, the instrumental
resolution and the fit may not be very sensitive to this width.  For example,
the GB880205 lines have FWHM of 3.0 and 13.4~keV where the resolution is 5.8
and 8.8~keV, and therefore the low energy line is unresolved while the high
energy line will be somewhat resolved. Unresolved lines can also be modeled by
a rectangular line profile where the flux is zero over an energy range (the
equivalent width) centered on the line centroid; this model requires only two
parameters. For the GB880205 lines we fit and compared the line significances
in simulations with two independent three parameter lines (six parameters in
total), two harmonically spaced three parameter lines (five parameters total),
two independent lines with rectangular profiles (four parameters) and two
harmonically spaced rectangular lines (three parameters). The issue is the
tension between the preference of statistical determinations of line
significance (e.g., the {\it F}-test) for models with fewer parameters and the
degradation in the fit when a spectrum is fit by an overly simplistic model.

\vfill\eject

\section{RESULTS}

We use the spectra from the two \Ginga\ detections to define the line types in
our calculations.  The first is based on the S1 segment of GB870303 (Graziani
et al. 1992); we use a new fit provided by E.~Fenimore (private communication,
1994),
\begin{equation}
N(E) = 5.76 \times 10^{-3} \left({E \over {100}}\right)^{-1.54}
   \exp \left[-42.5 \exp\left[-\left({{E-20.6}\over
   {2.5}}\right)^2\right]\right]
   \hbox{ ph-cm}^{-2}\hbox{-s}^{-1}\hbox{-keV}^{-1} ,
\end{equation}
where $E$ is in keV.  The line was observed to persist for 4s.  The second line
type models the lines in GB880205 (Murakami et al. 1988).  Again, we use a new
fit provided by E.~Fenimore (private communication, 1994),
\begin{eqnarray}
N(E) &=& N_C(E)- {{0.7037}\over{\sqrt{2\pi} 1.27}}
   \exp\left[ -{1\over2} \left({{E-19.4}\over{1.27}}\right)^2\right]
   \nonumber \\
   && -{{1.362}\over{\sqrt{2\pi}5.70}}
   \exp\left[-{1\over 2}\left({{E-38.8}\over{5.70}}\right)^2\right]
   \hbox{ ph-cm}^{-2}\hbox{-s}^{-1}\hbox{-keV}^{-1} \\
\hbox{where}\quad N_C(E) &=& 0.0819 (E/100)^{-0.872}
   \exp\left[-\left({{E}\over{250}}\right)\right]\quad , \quad
   E\le \hbox{282 keV} \nonumber \\
   &=& 0.0854 (E/100)^{-2} \quad , \quad E> \hbox{282 keV} \quad ; \nonumber
\end{eqnarray}
note that we have expressed the Gaussians in terms of ``standard deviations''
$\sigma$ (hence the numerical factors).  This line persisted 5.5s. The
continuum we use for GB880205 does not correspond to the best fit \Ginga\
models above 100~keV, but is instead the four parameter functional form
$N_C(E)$ (sometimes called the GRB model) which we find successfully describes
burst spectra (Band et al. 1993b).

For each line type we ran simulations at burst angles of $\theta=0^\circ,$
$15^\circ$, $30^\circ$, $46^\circ$, $60^\circ$ and 75$^\circ$ for a variety of
live times which were then translated into SNR/$\Delta E^{1/2}$; Figure~3 shows
the results of these simulations.  The spectra in GB870303 were fit over
$E=10-100$~keV (Fig.~3a) and $E=10-1200$~keV (Fig.~3b), while the two lines in
GB880205 were fit over $E=10-1200$~keV with a variety of line models
(Figs.~3c-f).  In addition, we ran simulations every 5$^\circ$ for the time the
features were observed to persist (4s for GB870303 and 5.5s for GB880205); the
resulting dependencies are shown by Figure~4.

The results of the simulations at a given burst angle were modeled empirically
by
\begin{equation}
p = \left[1-\tanh\left( {{s-s_0}\over{\Delta s}}\right)\right]/2
\end{equation}
where $p$ is the detection probability and $s$ is the SNR/$\Delta E^{1/2}$ (as
described in \S 3).  At $s=s_0$ we have $p=1/2$, while $\Delta s$ characterizes
the SNR/$\Delta E^{1/2}$ range over which $p$ changes from 0 to~1.  Here we use
an {\it F}-test probability of less than $10^{-4}$ as the detection criterion,
and the SNR/$\Delta E^{1/2}$ in the 25-35~keV band.  The resulting fits to the
simulations are shown by Figure~3 and listed in Table~1.  As can be seen, this
model works remarkably well.  It should be noted that these curves are fit to
a small number of empirical data points, each of which is the average of a
finite number of simulations.  Consequently there is an uncertainty in the
curves' true shape.  We assume there is actually a monotonic increase in
SNR/$\Delta E^{1/2}$ with burst angle to achieve a given detection probability,
yet the curves occasionally cross at small or large probabilities because of
these uncertainties.

Figure~3 and Table~1 establish that there is a significant dependence of the
detection probability not only on SNR/$\Delta E^{1/2}$ but also on burst angle.
Indeed the SNR/$\Delta E^{1/2}$ necessary for a detection increases
monotonically with burst angle; at small detection rates the empirical fits
cross in a few cases, most likely because the SNR/$\Delta E^{1/2}$ dependence
is undersampled, and the points were calculated with a finite number of
simulations. This dependence on burst angle can also be seen from Figure~4
where we show the detection probability for the two \Ginga\ lines with the
observed line persistence times.  We use detection criteria of {\it F}-test
probabilities less than $10^{-4}$ and $10^{-5}$.

As Figures~3c-f and the listings for GB880205 in Table~1 also show, reducing
the number of parameters by using a simpler line model does not change the
calculated detectability appreciably.  On the other hand, fitting the burst
spectrum over a broader energy range makes lines more detectable; thus a line
will be more significant when the spectrum is fit over a more extended energy
range, even when the continuum model requires more parameters.  To demonstrate
this point, we fit the same 200 realizations of the model for GB880205 in
eqn.~(4) (except that in this case the break energy in the exponential of the
GRB model $N_C$ is 300~keV, not 250) from 10 to 100, 300, 900 and 1200~keV with
different continuum models; the geometric mean of the $F$-test probabilities
are presented in Table~2. As can be seen, the fits become more significant as
the fitted spectrum is extended far above the high energy line at 38.8~keV.
Note that the GRB continuum model $N_C$ is ``correct'' since this is the model
used to create the spectra.  Yet the simpler, less ``correct'' continuum
models give more significant line
fits.  Adding lines to incorrect, overly simple continua not only fits the
lines but also compensates in part for the poor continuum model, and therefore
the improvement in the fit when the lines are added to the continuum is
greater, as is the apparent line significance.  Thus this greater line
significance is misleading.  Of course, we do not know the correct continuum
shape of real burst spectra; the GRB model is undoubtedly also too simple.
Thus our fits may also overestimate the significance of line candidates.

\section{DISCUSSION}

Using a heuristic model of a spectrum with a line we can develop some
approximate dependencies of $\langle \Delta \chi^2 \rangle$, the average
improvement in $\chi^2$ of the continuum+line model over the continuum model,
on various line parameters; these dependencies explain our results, and provide
guidance for future scalings (Ford et al. 1993).  Let the observed count
spectrum $f_{obs}$ be divided into $n$ channels of equal width $\Delta E$, with
the line falling in $n_L$ channels. Assume the observed continuum is at a
constant level $f_c$, and the line is at a level $f_L$ (less than $f_c$):  the
line profile in this simplified example is rectangular.  We now compare a
continuum model to a continuum+line model, where once again the continuum is
assumed to be at a constant flux, and the line has a rectangular profile.  On
average the best fit continuum+line model will be the same as the observed
spectrum without noise, while the continuum-only fit will be
$f_m=[(n-n_L)f_c+n_L f_L]/n$.

The noise has zero mean and non-zero variance $\sigma^2$.  Using model
variances and Poisson statistics we have for the $i$th channel $\sigma^2_i =
F_{M,i}/A\Delta t \Delta E$, where $F_{M,i}$ is the model flux averaged over
the channel, $A$ is the effective area of the detector, $\Delta t$ is the
live time, and $\Delta E$ is the channel width.  The observed spectrum
$f_{obs}$
is assumed to be the model spectrum $F_M$ with noise,
\begin{equation}
f_{obs,i} = F_{M,i}+\sigma_i \phi_i
\end{equation}
where $\phi_i$ is the normalized noise distribution:
$\langle \phi_i \rangle=0$ and $\langle \phi_i^2 \rangle=1$.  The
continuum+line model is ``correct'' and therefore
\begin{equation}
\langle \chi^2_{C+L} \rangle = \langle \sum_{i=1}^{n}
   {{\left(f_{obs,i}-F_{M,i}\right)^2} \over{\sigma_i^2}} \rangle
   = n
\end{equation}
while for the continuum-only model
\begin{eqnarray}
\langle \chi^2_C \rangle &=& \langle \sum_{i=1}^{n}
   {{\left(f_{obs,i}-F_{M,i}\right)^2} \over{\sigma_i^2}} \rangle \nonumber \\
   &=& n + {{A\Delta t \Delta E}\over{f_m}}
   \left[ \sum_{i=1}^{n-n_L} \left(f_c-f_m \right)^2
   + \sum_{i=1}^{n_L} \left(f_L-f_m \right)^2 \right] \\
   &=& n + {{A\Delta t \Delta E}\over{ (n-n_L)f_c +n_L f_L}} n_L (n-n_L)
   (f_c-f_L)^2 \quad . \nonumber
\end{eqnarray}
Identifying the observed line width as $\Delta E_{line}=n_L \Delta E$, the
equivalent width of the line as $eW = \Delta E_{line} (f_c-f_L)/f_c$, and the
SNR for a single continuum channel as
SNR$=[f_c A\Delta t \Delta E]^{1/2}$, we find
\begin{equation}
\langle \Delta \chi^2\rangle = \langle \chi_C^2 -\chi_{C+L}^2\rangle
   = {{(n-n_L)f_c}\over{(n-n_L)f_c+n_Lf_L}} \, (eW)^2
   \, {{(\hbox{SNR})^2}\over {\Delta E}} \, {1\over {\Delta E_{line}}} \quad .
\end{equation}
Note that since SNR$\propto \Delta E^{1/2}$, (SNR)$^2 /\Delta E $ is a function
of the average continuum flux and not the width of the energy range; indeed we
parameterize the line detectability by SNR/$\Delta E^{1/2}$.  Since our
heuristic calculation used the observed count spectrum and not the intrinsic
photon spectrum, $\Delta E_{line}$ is the observed line width which results
from the instrumental energy resolution and the intrinsic width.

As can be seen from eqn. (2), $F\propto \Delta \chi^2$, and we find that the
negative of the logarithm of the probability $P(>F)$ is approximately
proportional to $\Delta \chi^2$ for a large number of degrees-of-freedom.  Thus
the absolute value of the logarithm of the line significance is approximately
proportional to $\Delta \chi^2$ (i.e., $-\log P(>F) \propto \Delta \chi^2$).
In our case we are interested in the fraction of the distribution of $P(>F)$
which meets our detection criteria, and therefore the relationship between
detectability and $\Delta \chi^2$ is more complicated, but monotonic.  From our
simulations we see that the line detectability is dependent on the SNR/$\Delta
E^{1/2}$. In our simulations we fixed the line parameters, and therefore $eW$
and $\Delta E_{line}$ were not varied.  However, note that the line in GB870303
is detectable at lower SNR/$\Delta E^{1/2}$ than the two lines of GB880205
(compare Fig.~3b to Figs.~3c-f), which can be understood in terms of the
equivalent widths and line widths.  For the photon spectrum of the GB870303
line the $eW$ and $\Delta E_{line}$ are both approximately 6~keV.  The 19.4~keV
line of GB880205 has $eW=2.1$~keV and $\Delta E_{line}=3$~keV while the
38.8~keV line has $eW=7.3$~keV and $\Delta E_{line}=13$~keV.  However, the
analysis above deals with the observed count spectrum, not the photon spectrum.
Thus the GB880205 line at $19.4$~keV would be unresolved, and the 20.4~keV line
in GB870303 would be at best partially resolved, and thus both would be
observed to have the width of the detectors' resolution of about 6~keV.
Consequently the line in GB870303 would have a much greater equivalent width
than, and comparable observed width to, the $\sim20$~keV line in GB880205.
Similarly, the line in GB870303 would have a comparable equivalent width to,
but smaller observed width than, the $\sim40$~keV line in GB880205.  Thus the
dependencies in eqn.~(9) explain why the single line in GB870303 is
detectable at a lower SNR/$\Delta E^{1/2}$ than the lines in GB880205.

The simulations show that extending the energy range over which the spectrum is
fit increases the line significance, even if the continuum must be fit by a
more complicated continuum model.  This improvement in the significance results
from two effects.  First, increasing the number of degrees-of-freedom, $\nu_2$,
while holding $F$ fixed ($F$ is the statistic used for the $F$-test---eqn.~2)
decreases $P(\ge F)$ (increases the line significance); for a large number of
parameters (e.g., 6 for fitting GB880205) the factor of two increase in $\nu_2$
resulting from raising the upper cutoff $E_{max}$ from 100 to 1200~keV can
decrease $P(\ge F)$ by more than an order of magnitude.  Note that $F$ is
inversely proportional to the reduced $\chi_\nu^2=\chi^2_2/\nu_2$, which for a
good continuum+line fit will be approximately 1 regardless of the value of
$\nu_2$. Thus if $\Delta \chi^2$ is fixed, $F$ will most likely be fixed.
Second, including more continuum in
the fit forces a larger difference between
the continuum and continuum+line models, even when the added continuum is far
from the energy of the candidate line feature. As eqn.~(9) shows, $\Delta
\chi^2$ increases with the ratio of the continuum counts, $(n-n_L)f_c$, to the
counts in the line $n_L f_L$. The larger $\Delta \chi^2$ increases $F$ and
decreases $P(\ge F)$.  The first effect is in some sense an artifact of the
statistical method while the second effect results from more fully utilizing
the available spectral information.

We can calculate the effect on the significance of merely changing $\nu_2$.
First we find for each continuum type the $F_{1200}$ which gives the value of
$P_{\nu_2 \sim 215}(\ge F_{1200})$ calculated from the simulations for
$\nu_2\sim 215$ when $E_{max}=1200$~keV. Using this $F_{1200}$ we then find
$P_{\nu_2}(\ge F_{1200})$ for the values of $\nu_2$ applicable to the different
$E_{max}$.  Table~2 shows $\nu_2$ and the ratio $P(\ge F) / P_{\nu_2} (\ge
F_{1200})$ for the different combinations of continuum type and $E_{max}$.  As
can be seen, $P_{\nu_2}(\ge F_{1200})$ decreases less rapidly than $P(\ge F)$
as $\nu_2$ increases.  Note that the COMP model is poorly constrained by a
continuum up to only $E_{max}=100$~keV, while the GRB model is poorly
constrained for models to $E_{max}=100$ and 300~keV.

Given the above dependencies on equivalent width and SNR/$\Delta E^{1/2}$, the
dependence on burst angle may seem surprising.  However, the energy dependence
of the detector efficiency changes with angle, particularly at low energy where
the aluminum case and beryllium window have different absorption properties.
Between $\sim10$ and $\sim20$~keV the effective area is a maximum along the
detector normal, while at higher energies, where the aluminum is transparent,
the effective area is a maximum at a burst angle of $\sim 30^\circ$ (where the
NaI crystal's projected area is greatest).

Our simulations reveal a large distribution of fitted line parameters and
significances for the same input photon spectrum.  Figure~2 shows that even a
weak line, which would be considered a detection in only 15\% of all
realizations (with an {\it F}-test threshold of $10^{-4}$), can occasionally
appear to be extremely significant.  This distribution of line significances is
not related to fluctuations in the best fit line centroids, but clearly
increases with the equivalent width of the fitted line, as demonstrated by
Figure~5.  The wide dispersion in the fitted line centroid and equivalent width
are clear from Figures~5 and~6, respectively.  Consequently, the fitted
parameters to a line detection may not be an accurate description of the true
line.

An interesting result of our simulations is that the line detectability is
nearly the same whether the ``correct'' many-parameter line model or simpler
models with fewer parameters are used to fit the GB880205 simulated spectra. As
can be seen from the definition of the $F$ statistic (eqn.~[2]), the
significance of a line increases with a reduction in the number of parameters
(smaller $\Delta \nu$) or with an improvement of the fit (larger $\Delta
\chi^2$). The width of the 19.4~keV line is $\sim 2/3$ of the energy resolution
while the 38.8~keV line is $\sim 50\%$ larger than the resolution.  Thus the
fits are sensitive to the line profile, and eliminating the line width degrades
the fit sufficiently to balance the reduction in parameters.  Surprisingly,
fixing the ratio of the line centroids to be 2 does not increase the
detectability appreciably.  Note that reducing the number of line parameters
from 6 to 5 increases $F$ by only a factor of $\sim 1.2$ (since $\nu_2 \sim
200$, the difference in $\nu_2$ for the two models is only $\sim 0.5$\%) which
for line significances of order $10^{-4}$ increases the significance (i.e.,
decreases $P(>F)$) by a factor of only $\sim 2.7$.  Because of fluctuations the
best fit centroids are not at the input values, and therefore models with two
independent line centroids result in better fits than models where the lines
are constrained to be harmonic. The competition between simpler models and
better fits is always present in modeling observations; with our sequence of
models there is little preference for simplicity even when this simplicity
appears to be justified on physical grounds.  However, in other situations,
such as fitting lines which are truly unresolved, simpler models may result in
more significant fits.

While these detectability simulations were undertaken to support the comparison
of the BATSE and \Ginga\ observations, they also demonstrate BATSE's detection
capabilities.  As Figure~4 shows, the SDs could have detected the lines in
GB880205 up to angles of $\sim75^\circ$ while the line in GB870303 would have
been detected at small angles (less than $30^\circ$) in about a quarter of the
observations.  BATSE consists of an array of detectors, and a burst at any
point in the sky can be observed by a number of detectors at different burst
angles.  We therefore derived the average probability that the line would be
detected in at least one detector.

Thus we calculated the detector array's set of burst angles for 100,000
``bursts'' distributed randomly on the sky; these sets were binned by the
cosine of the burst angles. The probability of a detection by at least one
detector was computed for each set of burst angles using the detection
probabilities as a function of burst angle (e.g., based on the data in Fig.~3).
For most of the mission six SDs have been set at high gain (extending their
sensitivity to low energies), while a pair of opposite detectors have been
operated at low gain (effectively making them useless for line searches).
However, the low energy cutoff of one high gain detector is only $\sim 25$~keV
when operated at its highest gain; this detector is unable to detect lines at
$\sim20$~keV.  Only one detector of any pair of opposite detectors will have a
positive cosine which is necessary for the burst to fall in the detector's
field-of-view (unless both detectors have a cosine of zero).  We therefore
model the average detection probability using five detectors, of which four are
part of two detector pairs.  For comparison we also calculated the
probabilities for two, three and four detector pairs.

Of necessity we use a simplified model of the detector array in this
calculation. The true energy cutoffs may differ from 10 and 1200~keV since
without automatic gain control the gain of each detector (and therefore the
energy covered) drifts with time; also, various instrumental constraints cause
the gain to vary.  Next we assume that the data from all the SDs facing the
burst are available.  However, burst data are transmitted to Earth from only
four detectors based on their LAD count rates, which may be affected by Earth
scatter.  In addition, the LAD and SD axes are offset by $\sim18.5^\circ$.
Thus, data may be returned from one or two SDs with burst angles greater than
90$^\circ$ instead of from SDs with angles less than $90^\circ$.  Therefore, in
reality data from less than four burst-facing SDs may be returned to
Earth. Finally, Earth blockage was not included in the calculation since in
{\it CGRO} coordinates the Earth's position should average out over many {\it
CGRO} viewing periods.

Figure~7 shows the distribution of the cosine of the burst angle by detector
rank assuming 4, 5, 6 and 8 detectors covered the energy range relevant for
detecting lines.  The detectors are ranked by the order of their burst angles;
thus the first rank detector has the smallest angle, etc.  As expected,
operating more detectors at high gain results in a smaller average value of the
smallest burst angle. However, the relevant quantity is the detection
probability for the set of burst angles.  For the detector array as a whole a
given burst must be characterized by the persistence time and not by the
SNR/$\Delta E^{1/2}$ since a variety of SNR/$\Delta E^{1/2}$ values will result
for a given persistence time as a function of each detector's burst angle.

We calculated the average detection probability by averaging the net detection
probability of the 100,000 bursts in each of our detector arrays, as shown
by Fig.~8.  This net probability is the probability that the line feature would
be significant in at least one of the SDs; the probabilities for a single
detector at a given angle were used to calculate the net probability for the
entire detector array. Again as expected, the average detection probability
increases when more detectors cover the energy range of the line features.  The
observed persistence times of the two \Ginga\ bursts are indicated.  Thus the
line in GB870303 would be detected by BATSE with an average probability of
$\sim 1/8$, while the lines in GB880205 would almost always be detected. Note
that this calculation assumes the lines occur in spectra with the intensities
reported by \Ginga.  The line in GB870303 is detectable at longer persistence
times but at smaller values of SNR/$\Delta E^{1/2}$ than the lines in GB880205
because the continuum in the GB870303 spectrum is much weaker than for
GB880205, while the line equivalent width is greater.  Thus a line with the
equivalent width observed in GB870303 in a continuum as intense as GB880205
would be spectacular.

\section{SUMMARY}

By fitting simulated BATSE spectra of the two line sets observed by \Ginga\ we
find that BATSE should indeed be able to detect similar lines.  These
simulations provide us with the detectability as a function of a spectrum's
signal-to-noise ratio, a relationship necessary for calculations of the number
of lines which would have been detected if present in the BATSE bursts. We note
that these are calculated instrumental capabilities.

These simulations have also provide guidance into the analysis of line
candidates.  We find that fitting spectra over a broad energy range increases
the significance of line candidates.  Decreasing the number of parameters by
using physically motivated simplified models (e.g., assuming multiple lines are
harmonically spaced or eliminating the intrinsic line width because it is
smaller than, or comparable to, the instrumental resolution) may not result in
the expected increase in significance, as we found in simulations with the pair
of lines in GB880205.  These lines have widths comparable to the instrument
resolution, and therefore it is not surprising that simplifying the profile
degrades the fit.  Although the lines were harmonically spaced in the photon
spectrum, forcing the fitted lines to be harmonic did not significantly improve
the line detectability, presumably because fluctuations shift the best fit line
centroids.  We conclude that the degradation in the quality of the fit
counterbalances the reduction in the number of line parameters.

Our simulations demonstrate that BATSE is able to detect the line in GB870303
in about an eighth of the possible realizations and positions on the sky
relative to {\it CGRO}, while the lines in GB880205 are almost always
detectable.  For most of the mission five of the eight spectroscopy detectors
have been operated at high gain to maximize the sensitivity to absorption
lines; increasing the gain of two more detectors (one detector cannot be pushed
to high enough gain) will increase the average detection probabilities but not
by a large enough factor to qualitatively change BATSE's line detection
abilities.

The implications of these line detectabilities for issues such as consistency
between the BATSE and \Ginga\ observations will be explored in future papers of
this series.

\acknowledgments{We thank E.~Fenimore and P.~Freeman for information regarding
the \Ginga\ observations.  We also thank the referee for pointing out the
apparent improvement in line significance which results from the increase in
the
number of degrees-of-freedom as the fitted energy range is extended. The work
of the UCSD group is supported by NASA contract NAS8-36081.}

\vfill\eject\addtocounter{page}{2}

\vfill\eject

\centerline{\bf FIGURES}
\bigskip
Fig. 1.---Simulated count spectra of (a) the S1 line in GB870303 and (b) the
lines in GB880205 as they might have appeared in the BATSE SDs.  The
curve is the photon spectrum convolved with the detector response at a
burst angle of $30^\circ$.  The data points, rebinned into bins
half a resolution element wide, are a possible realization of this count
spectrum assuming Poisson statistics.

Fig. 2.---Distribution of $F$-test probabilities.  Based on 2400 simulations of
the S1 line of GB870303 which persisted 4s, observed at a burst angle of
15$^\circ$ and fit over $E=10-100$~keV.  The two vertical lines indicate the
detection thresholds of $F$-test probabilities less than $10^{-4}$ (short
dashes) and $10^{-5}$ (long dashes); the fraction of the distribution to the
left of these lines is the detection probability.

Fig. 3.---Detection probability vs. SNR/$\Delta E^{1/2}$ in the 25-35~keV band
for a variety of burst angles.  From left to right, the angles are 0$^\circ$,
15$^\circ$, 30$^\circ$, 46$^\circ$, 60$^\circ$ and 75$^\circ$.  An $F$-test
probability less than $10^{-4}$ was used as the detection criterion.  The data
points are the results of the simulations and the solid curves are empirical
fits to these points.  Panels (a) and (b) are for the S1 segment of GB870303
fit over $E=10-100$~keV and $E=10-1200$~keV, respectively.  Panels (c)--(f)
show the detectability of the GB880205 line over $E=10-1200$~keV where: (c)
rectangular line profiles are used and the two line centroids are required to
be harmonic, $E_2=2E_1$ (3 line parameters); (d) rectangular line profiles are
used and the two line centroids are fit independently (4 parameters); (e) the
three parameter line model is used with harmonic lines (5 parameters); and (f)
the three parameter line model is used with the two centroids fit independently
(6 parameters).

Fig. 4.---Detection probability vs. burst angle for the two \Ginga\ lines at
the observed persistence times.  The upper data points are for an $F$-test
probability less than $10^{-4}$ and the bottom set of points for $10^{-5}$. The
panels correspond to the same cases as in Fig.~3.

Fig. 5.---Fitted equivalent width vs. $F$-test probability for a set of 2400
simulations of the S1 line of GB870303 which persisted 4s at a burst angle of
15$^\circ$. The fitted equivalent width has been normalized to the value for
the input line.

Fig. 6.---Distribution of fitted line centroid energies for a set of 2400
simulations of the S1 line of GB870303 which persisted 4s at a burst angle of
15$^\circ$ fitted over $E=10-100$~keV.  The input line centroid is indicated
by the vertical dashed line. The fitted energy was constrained to fall between
15 and 30~keV.

Fig. 7.---Distribution of the cosine of the burst angle by detector rank.  The
panels show the distribution for a)~two detector pairs, b)~5 detectors
including two pairs, c)~three pairs, and d)~four pairs.  The detectors are
ranked by the size of their burst angles.  For a detector to be included the
burst must occur in its field-of-view (non-negative cosine), and the gain must
be high enough to observe lines at $\sim20$~keV.

Fig. 8---Average detection probability vs. the time a feature persisted for
(a)~the S1 line of GB870303 and (b)~the lines in GB880205.  From top to bottom,
the curves are for four detector pairs (small dashes), three pairs
(dash-dot-dot), 5~detectors including two pairs (solid), and two pairs (long
dashes).  The vertical lines indicate the observed persistence times. The burst
intensities reported by \Ginga\ are used.

\end{document}